\documentclass{mn2e}
\usepackage{epsfig}

\newif\ifAMStwofonts
\newcommand{\novamon}{A0620--00}
\newcommand{\novamus}{XN Mus 1991}
\newcommand{\novavel}{XN Vel 1993}
\newcommand{\novamuslong}{X-ray Nova Muscae 1991}
\newcommand{\novavellong}{X-ray Nova Velorum 1993}
\title[Fast Photometry of Quiescent Soft X-ray Transients] 
{Fast Photometry of Quiescent Soft X-ray Transients with the 
Gemini-South Acquisition Camera}

\author[R. I. Hynes et al.]
       {R. I. Hynes$^{1,2}$\thanks{E-mail: rih@astro.as.utexas.edu;
       Hubble Fellow}, 
	P. A. Charles$^1$, 
        J. Casares$^3$,
	C. A. Haswell$^4$, 
	C. Zurita$^3$,
	T. Shahbaz$^3$ \\
$^1$Department of Physics and Astronomy, University of Southampton, 
    Southampton, SO17 1BJ\\
$^2$Astronomy Department, The University of Texas at Austin, 1
       University Station C1400, Austin, Texas 78712-0259, USA\\
$^3$Instituto de Astrof\'\i{}sica de Canarias, 38200 La Laguna,
    Tenerife, Spain\\
$^4$Department of Physics and Astronomy, The Open University, Walton
    Hall, Milton Keynes, MK7 6AA} 
\date{Accepted 2002 November 22.
      Received 2002 October 31;
      in original form 2002 October 31}

\pagerange{\pageref{firstpage}--\pageref{lastpage}}
\pubyear{2002}

\begin{document}
\maketitle
%
%%%%%%%%%%%%%%%%%%%%%%%%%%%%%%%%%%%%%%%%%%%%%%%%%%%%%%%%%%%%%%%%%%%%%%%%%%%%%%%
%
\begin{abstract}
We present a compilation of high time-resolution photometric
observations of quiescent soft X-ray transients obtained with the
acquisition camera of Gemini-South.  \novamon\ was observed with a
short cycle time and high precision.  Superimposed on the ellipsoidal
modulation we find several prominent flares together with weaker
continual variability.  The flares seen sample shorter timescale than
those reported in previous observations, with rise times as low as
30\,s or less; most flares show unresolved peaks.  The power density
spectrum (PDS) of \novamon\ appears to exhibit band-limited noise
closely resembling the X-ray PDS of black hole candidates in their low
states, but with the low-frequency break at a lower frequency.
\novamuslong\ shows much larger amplitude flares than \novamon\ and if
a break is present it is at a lower frequency.  \novavel\ shows very
little flaring and is, like \novamon, dominated by the ellipsoidal
modulation.  We discuss the possible origins for the flares.  They are
clearly associated with the accretion flow rather than an active
companion, but whether they originate in the outer disc, or are driven
by events in the inner region is not yet resolved.  The similarities
of the PDS to those of low/hard state sources would support the latter
interpretation, and the low break frequency is as would be expected if
this frequency approximately scales with the size of an inner
evaporated region.  We also report the discovery of a new variable
star only 14\,arcsec from \novamus.  This appears to be a W~UMa star,
with an orbital period of about 6\,hrs.
\end{abstract}
%
%%%%%%%%%%%%%%%%%%%%%%%%%%%%%%%%%%%%%%%%%%%%%%%%%%%%%%%%%%%%%%%%%%%%%%%%%%%%%%%
%
\begin{keywords}
accretion, accretion discs -- binaries: close -- stars: individual:
V616~Mon, GU~Mus, MM~Vel
\end{keywords}
%
%%%%%%%%%%%%%%%%%%%%%%%%%%%%%%%%%%%%%%%%%%%%%%%%%%%%%%%%%%%%%%%%%%%%%%%%%%%%%%%
%
\section{Introduction}
\label{IntroSection}
Soft X-ray transients (SXTs), also referred to as X-ray novae and
black hole X-ray transients, are low-mass X-ray binaries in which long
periods of quiescence, typically decades, are punctuated by very
dramatic X-ray and optical outbursts, often accompanied by radio
activity (Tanaka \& Shibazaki 1996; Cherepashchuk 2000).  In outburst
a number of X-ray spectral states are seen, most commonly the
high/soft state and the low/hard state.  An intermediate and very high
state have also been identified.  In the high/soft state, X-ray
emission is dominated by thermal emission from an accretion disc
extending to close to the last stable orbit around a black hole.  In
the low/hard state, the inner disc is believed to be truncated and
emission appears to originate from an extended corona.  Direct support
for this picture is provided by the low/hard state source,
XTE~J1118+480 (Hynes et al.\ 2000), in which an inner disc radius of
at least 50\,R$_{\rm Sch}$, and probably $\sim350$\,R$_{\rm Sch}$, is
required (McClintock et al.\ 2001; Chaty et al.\ 2002).  A similar
scenario is advanced by various advective models for the quiescent
state (see Narayan, Garcia \& McClintock 2001 and references therein),
but with the disc truncated at larger radii, $10^3$--$10^5$\,R$_{\rm
Sch}$.  Attempts have been made to unify these spectral states within
the advective picture (Esin et al.\ 1997), and Esin et al.\ (2001) did
achieve some success in fitting the broad band spectrum of the
low/hard state source XTE~J1118+480 with an advective model.

The states of SXTs are classified by their X-ray timing properties as
well as by their spectra.  With the exception of the quiescent state,
these have been well studied (e.g.\ van der Klis 1995; Wijnands \& van
der Klis 1999).  The high/soft state shows a low level of red noise,
with no detected low frequency break.  The low/hard state and very
high state exhibit a higher level of band-limited noise, with a
low-frequency break at $\sim$0.02--30\,Hz, and sometimes superposed
QPOs.  Band-limited noise also appears to be seen in the Seyfert 1
galaxy NGC~3516 (Edelson \& Nandra 1999), but with a much lower
cutoff, $4\times10^{-7}$\,Hz.  This suggests an approximate scaling
with black hole mass, and presumably with the scale of the accretion
region.  The AGN data do not extend to low enough frequencies to be
confident that the break is analogous to the low frequency break in
the low/hard state, however; it could actually correspond to a higher
frequency turnover (Uttley 2002, priv.\ comm.).

Similar properties might be expected for quiescent SXTs, as the
structure of the flow is believed to be similar to that in the
low/hard state.  Observations of quiescent state variability are much
more difficult, however.  Sub-orbital variability is known to be
present at X-ray energies in the brightest source, V404~Cyg (Wagner et
al.\ 1994; Kong et al.\ 2002), but even this is faint; only
$\sim0.15$\,photons\,s$^{-1}$ are detected with {\it Chandra}.
Variability can be more effectively studied in the optical where
reasonable count rates are possible, and several photometric (Haswell
et al.\ 1992; Pavlenko et al.\ 1996; Zurita, Casares \& Shahbaz 2002a)
and spectrophotometric (Hynes et al.\ 2002) studies have been
performed.  The origin of the variability remains uncertain, however,
with plausible possibilities including direct optical emission from an
advective region, reprocessed X-ray variability, magnetic reconnection
events in the disc, and flickering from the accretion stream impact
point.  It may be that a combination of these factors are important on
different timescales, and that not all objects are dominated by the
same source of variability.  It is therefore important to perform a
comparative study of the class as a whole to determine if there is
just one type of variability or two or more with distinct
characteristics.  By doing this we can hope to isolate the
contribution, if any, from variations in the inner flow and hence
probe its nature.

Even in the optical, count rates from existing data are quite low,
limiting the time-resolution achieved, and severely compromising data
quality for fainter objects.  As we are studying aperiodic
variability, the only good solution to this problem is to increase the
count rate with a larger aperture telescope.  Consequently we have
embarked upon a survey of fast variability with the Acquisition Camera
on Gemini-South.  Our main goals are to explore the bright targets on
faster timescales than previously possible, and to study variability
effectively even in faint objects.  We present here high time
resolution data on the bright prototypical SXT, \novamon\ (=V616~Mon),
together with lower resolution lightcurves of the fainter objects
\novamuslong\ (GU~Mus), and \novavellong\ (MM~Vel).

%
%%%%%%%%%%%%%%%%%%%%%%%%%%%%%%%%%%%%%%%%%%%%%%%%%%%%%%%%%%%%%%%%%%%%%%%%%%%%%%%
%
\section{Observations}
\label{DataSection}
Photometric observations of the three quiescent SXTs were obtained in
service mode with the Acquisition Camera (AcqCam) on Gemini-South on
2001 December 15 and 2002 January 11 and 15.  A $V$ filter was used
with exposures between 6 and 60\,s.  The minimum dead-time was
primarily dictated by the CCD readout time, 1.7\,s, but a variable
additional delay was involved in data transfer.  The typical actual
dead time between exposures was about 2.2\,s.  Full details are given
in Table~\ref{ObsTable}.  Routine bias and dark subtraction and
flat-fielding corrections were applied to all the data before
distribution and appear satisfactory.

\begin{table*}
\caption{Log of Gemini-South AcqCam observations.}
\label{ObsTable}
\begin{center}
\begin{tabular}{lllrlll}
\hline
Object & Date & UT range & Exposures & Airmass & Seeing & $\left<V\right>$ \\
\noalign{\smallskip}
\novamon & 2001 Dec 15 & 02:26--06:30 & $1705\times6$\,s 
& 1.2--1.7 & 0.6--1.0 & 18.1 \\
\novamus & 2002 Jan 11 & 04:32--08:35 & $219\times60$\,s 
& 1.3--1.6 & 0.5--1.1 & 20.3 \\
\novavel & 2002 Jan 15 & 04:23--08:15 & $425\times30$\,s 
& 1.0--1.3 & 0.5--0.7 & 21.7 \\
\noalign{\smallskip}
\hline
\end{tabular}
\end{center}
\end{table*}

\novamon\ and \novamus\ were relatively bright compared to the
background and are not crowded (no significant contaminating stars
closer than 5\,arcsec.) These were therefore straightforward to
analyse.  Lightcurves of each object and several comparison stars were
extracted using aperture photometry.  For \novamon, a weighted mean of
the magnitudes of two nearby brighter comparisons was used for
differential photometry; for \novamus\ four brighter comparisons were
used.  In each case, two fainter nearby non-variable comparisons of
the same brightness as the target were used to check the photometric
accuracy.  The photometric apertures (0.7\,arcsec for \novamon, and
0.6\,arcsec for \novamus) were chosen to minimise the variance in the
lightcurve of one of the faint comparison stars.  From these
comparisons we estimate 1\,$\sigma$ accuracies of 0.8\,percent per
exposure for the flux lightcurve of \novamon\ and 1.7\,percent for
\novamus.

\novavel\ was more problematic as it lies close to a brighter star.
Fortunately, the seeing was good; 0.5--0.7\,arcsec FWHM for most of
the run with a few images at the beginning with FWHM up to
0.9\,arcsec.  An additional complication was that the immediate field
of \novavel\ was located close to the seam between the two halves of
the detector.  To minimise the effect of this on sky estimates, we
interpolated over the gap.  \novavel\ itself was not affected and we
did not use any stars that were on the seam either as PSF or
comparison stars.  Photometry of \novavel\ and a number of comparison
stars was obtained by PSF fitting using the {\sc iraf}\footnote{IRAF
is distributed by the National Optical Astronomy Observatories, which
are operated by the Association of Universities for Research in
Astronomy, Inc., under cooperative agreement with the National Science
Foundation.} implementation of {\sc daophot}.  PSF fitting with AcqCam
data is difficult, as the camera optics yield a position dependant
PSF.  This can largely be dealt with by choosing many PSF stars
surrounding the objects of interest and modelling position dependence.
We restricted the analysis to the immediate region of \novavel, with 7
suitable PSF stars, and allowed only linear variations (with respect
to $x$ and $y$) in the PSF; there were too few PSF stars to use a
higher order model.  This process did leave small residuals in the
core of the PSF of the brighter stars, but the wings are well
subtracted.  Fortunately, in most of our images, \novavel\ is
sufficiently resolved from the brighter star that the latter should
not be a significant problem.  For the images at the beginning of the
run, however, the poorer seeing caused problems and these images were
excluded from the extracted lightcurve.  Excluding these points, the
extracted lightcurves were insensitive to whether a variable PSF was
used or not, and to the adopted fitting radius.  Consequently we
believe that the lightcurve of \novavel\ is not significantly
contaminated by difficulties in the fitting.  Differential photometry
was performed with respect to six nearby comparison stars, all
brighter than \novavel\ and relatively isolated.  Another comparison
star of similar brightness to \novavel\ was extracted.  Both the
formal errors and the scatter in the comparison lightcurve give a
1\,$\sigma$ uncertainty of 0.9\,percent for this star.  Since
\novavel\ is blended, it is subject to larger errors; the formal error
estimate is 1.1\,percent.

For all three objects, an approximate photometric calibration was
applied with respect to several standard fields observed at low
airmass on the same night.  Colour terms were neglected as the targets
were observed in a single band only.
%
%%%%%%%%%%%%%%%%%%%%%%%%%%%%%%%%%%%%%%%%%%%%%%%%%%%%%%%%%%%%%%%%%%%%%%%%%%%%%%%
%
\section{Lightcurves}

\subsection{\novamon}
\label{NovaMonLCSection}
\begin{figure}
\begin{center}
\hspace*{-8mm}\epsfig{angle=90,width=3.8in,file=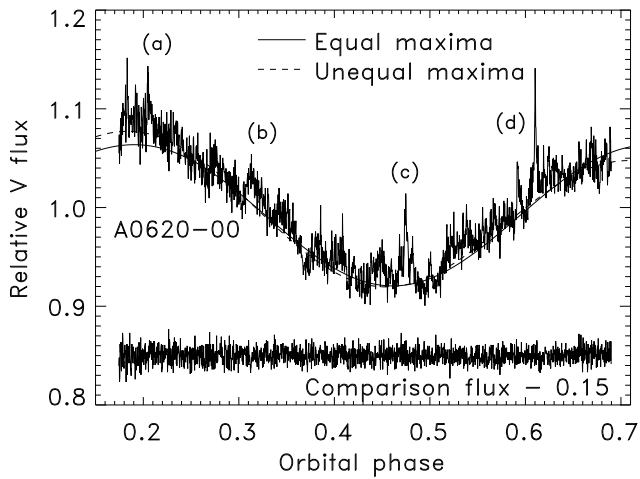}
\caption{Lightcurve of \novamon.  The abscissa is orbital phase with
respect to the ephemeris of Leibowitz et al.\ (1998).  A small,
$\sim0.05$, phase offset is present.  The comparison star has the same
average brightness as \novamon, but has been offset downwards by 0.15
units for clarity.  The fits to the ellipsoidal modulation are based
on a double sinusoid (fundamental on the orbital period plus first
harmonic).  In the equal maxima case, the phases of the sinusoids have
been fixed to produce a lightcurve like a pure ellipsoidal case.  In
the unequal case, the phasing is allowed to float to better fit the
data, as appropriate if the orbital lightcurve is distorted by other
effects.  Letters indicate the regions expanded in
Fig.~\ref{FlareFig}.}
\label{NovaMonLCFig}
\end{center}
\end{figure}

The lightcurve of \novamon\ is shown in Fig.~\ref{NovaMonLCFig}.  It
is clearly dominated by the ellipsoidal modulation due to the
distortion of the companion star (c.f.\ McClintock \& Remillard 1986).
Since this is not associated with accretion variability we should
remove this to isolate the flares.  We do this in a comparable way to
Zurita et al.\ (2002a), by approximating it with two sine waves at the
orbital frequency and its first harmonic.  Most of the modulation is
at the first harmonic but the fundamental provides for a variation in
the minima and/or maxima.  For an ideal pure ellipsoidal modulation
the maxima are equal, but the minima differ.  In practise, lightcurves
of quiescent SXTs in general, and \novamon\ in particular, often
exhibit unequal maxima as well (e.g.\ Haswell 1996; Leibowitz, Hemar
\& Orio 1998).  This may be due to distortion of the orbital
lightcurve by light from the stream impact point, starspots, and/or
persistent superhumps; the latter definitely appear to be seen in one
source, XTE~J1118+480, albeit in the last stages of outburst decline
rather than true quiescence (Zurita et al.\ 2002b).  We consider both
cases; for the pure ellipsoidal lightcurve we fix the relative phases
of the sine waves to produce equal maxima.  We also allow the phases
to vary independently.  We fit using an iterative rejection scheme to
approximately fit the lower envelope of the lightcurve, by rejecting
points more than 2\,$\sigma$ above the fit, then refitting.  This is
repeated until no new points are rejected.  We show the results of the
fitting in Fig.~\ref{NovaMonLCFig}.  Allowing the maxima to differ
does improve the fit somewhat, but the difference is not dramatic.

\begin{figure}
\begin{center}
\hspace*{-5mm}\epsfig{angle=90,width=3.6in,file=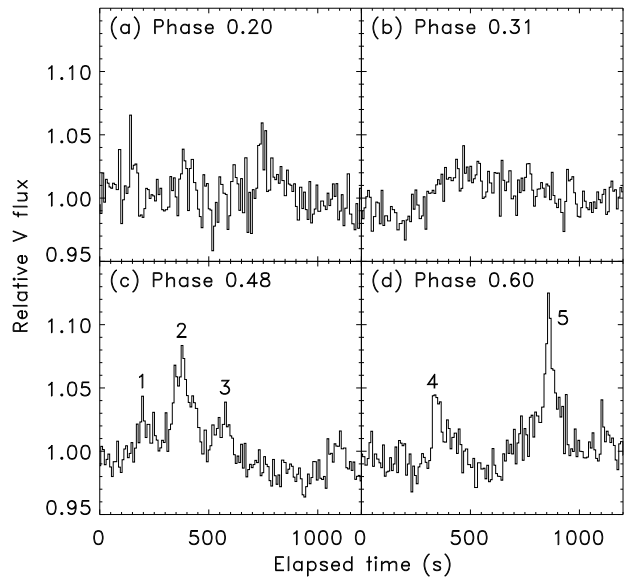}
\caption{Close-up view of a selection of flares in \novamon.  The
ellipsoidal modulation has been subtracted and replaced by its average
value.  Time and flux scales are the same for easy comparison of flare
amplitudes and durations.  The zero points of elapsed time are
arbitrary.  Numbered flares have their properties summarised in
Table~\ref{FlareTimescaleTable}.}
\label{FlareFig}
\end{center}
\end{figure}

Superposed on the ellipsoidal modulation are many rapid flares; the
strongest are shown in Fig.~\ref{FlareFig}.  Similar flares in
\novamon\ have been reported by Haswell (1992) and Zurita et al.\
(2002a).  We sample events of shorter duration, however; both of the
previous studies had a time-resolution of 30--40\,s, which would
barely have resolved our shortest events.  For the most prominent and
distinct flares shown in Fig.~\ref{FlareFig}, we have estimated some
characteristics of each flare; the peak amplitude, equivalent duration
(c.f.\ Zurita et al.\ 2002a), and rise and decay e-folding timescales,
based on an exponential fit.  These are not a representative sample,
as they are selected to be the most extreme, best defined events.  It
can be seen that these flares lie at the extreme-low end of the
distribution of equivalent durations presented by Zurita et al.\
(2002a).  E-folding timescales are typically 30--80\,s, although one
event (flare 5) rises much more rapidly than this.  There is no
consistent asymmetry to the flares, although individual events may be
asymmetric; again the most extreme behaviour is shown by flare 5.  We
have also characterised the activity level of the lightcurve using
similar nomenclature to Zurita et al.\ (2002a) in
Table~\ref{FlareTable}.

\begin{table}
\caption{Properties of selected individual flares from \novamon.  The
  peak is defined as a fraction of the mean level.  The equivalent
  duration is the total counts in the flare divided by the mean
  counts.  The rise and decay times are e-folding times from
  exponential fits to the rising and decaying segments of the flare profile.}
\label{FlareTimescaleTable}
\begin{center}
\begin{tabular}{lrrrr}
\hline
\noalign{\smallskip}
Flare & Peak & Equivalent   & Rise     & Decay \\
No.   &      & duration (s) & time (s) & time (s) \\
\noalign{\smallskip}
1 & 0.05 & 2.8 & 30 & 76 \\
2 & 0.10 & 9.5 & 29 & 69 \\
3 & 0.05 & 4.0 & 80 & 39 \\
4 & 0.06 & 6.2 &  7 & 81 \\
5 & 0.12 & 7.7 & 34 & 32 \\
\noalign{\smallskip}
\hline
\end{tabular}
\end{center}
\end{table}

\begin{table*}
\caption{Properties of Gemini lightcurves.  $v_{\rm obs}$ is the
spectroscopic veiling.  $v_d'$ is the contribution due to the {\em
  non-variable} disc light.  $\overline{z}_f$  is the mean flare
flux and $\sigma_z$ its standard deviation, both expressed as a
fraction of the mean flux.  $\sigma_z^*=\sigma_z / v_d'$ and $\eta$ is
the fraction of the average veiling due to the flares.}
\label{FlareTable}
\begin{center}
\begin{tabular}{lllllll}
\hline
Object & $v_{\rm obs}$ & $v_d'$ &$\overline{z}_f$ & $\sigma_z$ & $\sigma_z^*$
& $\eta$ \\
\noalign{\smallskip}
\novamon & $6\pm3$ (H$\alpha$), $17\pm3$ (H$\beta$) 
& $12\pm3$ & 0.009 & 0.015 & 0.12 & 0.07 \\
%         & $17\pm3$ (H$\beta$)  &          &       &       &    &    \\
\novamus & $\sim54$ (5000\,\AA), $15\pm6$ (6400\,\AA) 
& $26\pm4$ & 0.23  & 0.14 & 0.56 & 0.47 \\
%         & $15\pm6$ (6400\,\AA) &          &       &       &    &    \\
\novavel & $65\pm5$ (6300\,\AA) & $64\pm5$ & 0.024 & 0.023 & 0.04 & 0.04 \\
\noalign{\smallskip}
\hline
\end{tabular}
\end{center}
\end{table*}

\subsection{X-ray Nova Mus 1991}

\begin{figure}
\hspace*{-8mm}
\epsfig{angle=90,width=3.8in,file=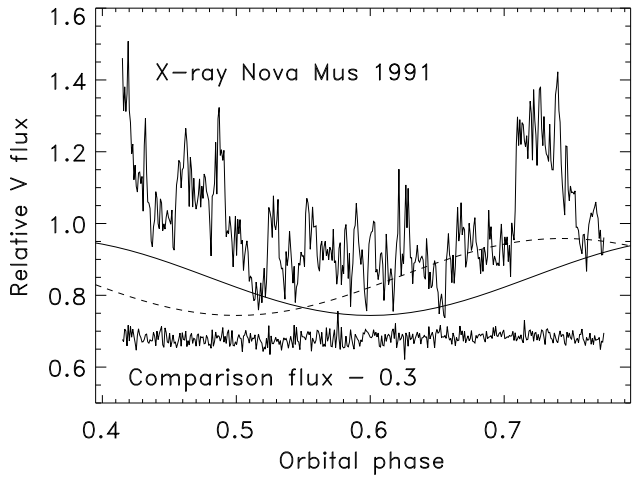}
\caption{Lightcurve of \novamus.  The abscissa is orbital phase with
respect to the ephemeris given in Shahbaz, Naylor \& Charles (1997).  The
comparison star has the same average brightness as \novamus, but has
been offset downwards by 0.3 units for clarity.  
The dashed line indicates the expected ellipsoidal modulation.  This
is not consistent with the data, but allowing an 0.12 phase offset
(solid line) improves the situation.
}
\label{NovaMusLCFig}
\end{figure}

The lightcurve of \novamus\ is shown in Fig.~\ref{NovaMusLCFig}.
Large amplitude aperiodic variability is dominant, to the extent that
any contribution from an ellipsoidal variation is not obvious in these
data, and cannot be fitted.  Ellipsoidal modulations have previously
been reported with a full amplitude in $B+V$ of $\sim$0.2--0.35\,mag
(Remillard, McClintock \& Bailyn 1992; Orosz et al.\ 1996).  Their
apparent absence is probably due to a combination of large flare
amplitude, small ellipsoidal amplitude and relatively long orbital
period.  We show in Fig.~\ref{NovaMusLCFig} the expected modulation,
assuming the ephemeris of Shahbaz et al.\ (1997) and a full amplitude
of 0.27\,mag.  It clearly is not consistent with the data.  However,
there may be an error in extrapolating the ephemerides over several
years, as appeared to be the case for \novamon; using the formal error
estimate of Shahbaz et al.\ (1997) this corresponds to an uncertainty
of 0.06 in phase at our epoch.  Fig.~\ref{NovaMusLCFig} also shows the
effect of including a phase offset of 0.12 (i.e.\ 2\,$\sigma$), which
is not unreasonable; the agreement with the lower envelope of the data
is now acceptable, and with this allowed for our data are consistent
with an ellipsoidal amplitude comparable to that previously reported.
There are too few points defining the lower envelope to reliably fit a
model with both the phase and amplitude variable, however, so we
neglect the relatively small ellipsoidal contribution in subsequent
sections unless explicitly noted.  Since the flaring in \novamus\ is
of such large amplitude, relative to the ellipsoidal contribution,
this should not introduce a large error.

Since the time-resolution was much less than for \novamon, we cannot
study rise and decay timescales of flares so readily although
relatively fast timescales are clearly present.  For example the large
and extended flare at the end of the observation represents a net
increase of about 40\,percent, with an e-folding time of $\sim50$\,s.
As for \novamon, the activity level of the lightcurve is characterised
in Table~\ref{FlareTable} to allow comparison with the results of
Zurita et al.\ (2002a).

\subsection{X-ray Nova Vel 1993}

\begin{figure}
\hspace*{-8mm}
\epsfig{angle=90,width=3.8in,file=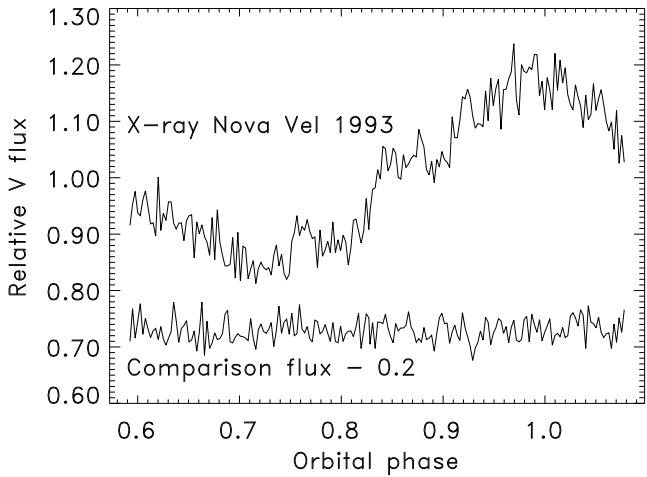}
\caption{Lightcurve of \novavel.  The abscissa is orbital phase with
respect to the updated ephemeris derived by Gelino (priv.\ comm.).
The comparison star has almost the same average brightness as
\novavel, but has been offset downwards by 0.3 units for clarity.  The
fits to the ellipsoidal modulation are based on a double sinusoid
model exactly as described for \novamon.}
\label{NovaVelLCFig}
\end{figure}

The lightcurve of \novavel\ is shown in Fig.~\ref{NovaVelLCFig} phased
on the new ephemeris of Gelino (priv.\ comm.).  This is dominated by
an apparent ellipsoidal modulation (c.f.\ Shahbaz et al.\ 1996).
Superposed on the ellipsoidal modulation do appear to be some flares.
The flare amplitude is much less than in \novamus, but is comparable
to that seen in \novamon.  Parameters of the detrended variability are
summarised in Table~\ref{FlareTable}.

%%%%%%%%%%%%%%%%%%%%%%%%%%%%%%%%%%%%%%%%%%%%%%%%%%%%%%%%%%%%%%%%%%%%%%%%%%%%%%%
%
\section{Power density spectra}
\subsection{\novamon}

To quantify the range of timescales present, we calculate a power
density spectrum (PDS), after removing the fitted ellipsoidal
modulation from the lightcurve.  Since the sampling was not perfectly
uniform we calculate a Lomb-Scargle periodogram (see Press et al.\
1992 and references therein) and normalise it in the same way as is
common for Fourier transform PDS.  Following the suggestion of
Papadakis \& Lawrence (1993), we bin and fit the PDS in logarithmic
space, i.e.\ each bin is evaluated as $\left< \log p \right>$ rather
than $\log \left<p\right>$ and we then fit to the values of
$\left<\log p \right>$.  Errors on the binned logarithmic power are
estimated from the standard deviation of points within the bin.  White
noise has been estimated and subtracted by fitting the highest
frequencies with a white noise plus red noise model, but the white
noise was not large, since the photometric precision was good.

\begin{figure}
\begin{center}
\hspace*{-8mm}
\epsfig{angle=90,width=3.8in,file=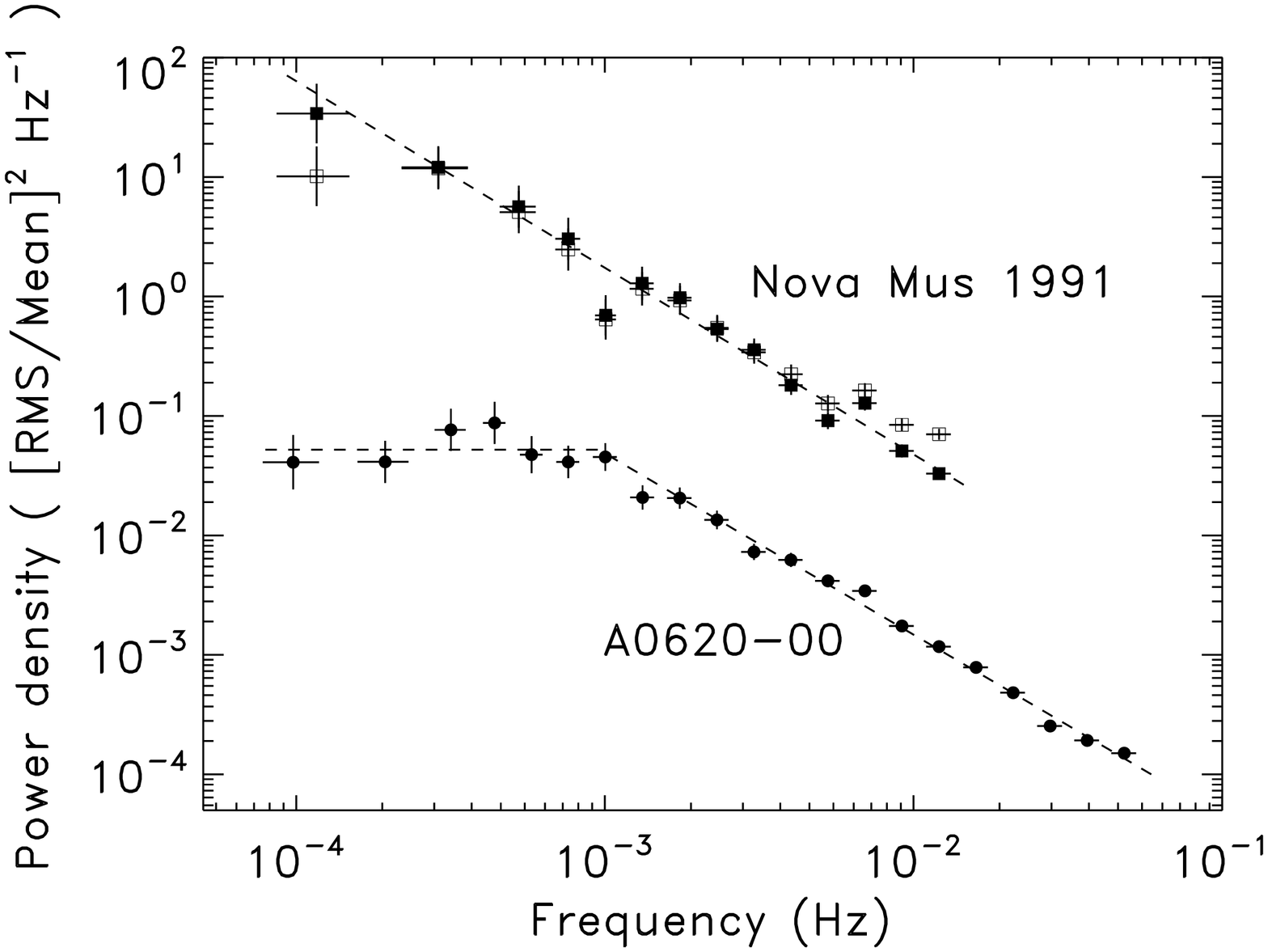}
\caption{PDS of \novamon\ and \novamus.  The white noise component,
fitted at the highest frequencies, has been subtracted.  The fitted
red noise slopes are $-1.52$ for \novamon\ and $-1.56$ for \novamus.
The break frequency in \novamon\ is at $9.5\times 10^{-4}$\,Hz,
corresponding to a timescale of $\sim20$\,min.  For \novamus, filled
squares indicate the PDS of the raw lightcurve, the open squares
indicate the effect of removed the representative ellipsoidal
modulation shown in Fig.~\ref{NovaMusLCFig}.}
\label{CompPDSFig}
\end{center}
\end{figure}

The derived PDS is shown in Fig.~\ref{CompPDSFig}.  At higher
frequencies we see a well defined power-law.  Below 1\,mHz (i.e.\
(20\,min)$^{-1}$), this appears to flatten.  The overall broken
power-law form is strikingly similar to the X-ray PDS of SXTs in the
low/hard state, as will be discussed in
Section~\ref{PDSDiscussionSection}.  To characterise it numerically,
we fitted it with a model which has a red noise power-law above a
break frequency and is flat below that.  We derive a power-law slope
of $-1.52$, similar to low/hard state SXTs and a break frequency at
$9.5\times 10^{-4}$\,Hz.

There are a number of potential pitfalls with this analysis.  In
detrending the ellipsoidal modulation we may have removed low
frequency accretion variability as well and hence flattened the PDS
artificially.  Also, with only 4\,hrs of data the shape of the
low-frequency PDS will be subject to fluctuations due to the
individual realisations of the spectrum.  Finally there may be some
aliasing problems at high frequencies due to the deadtime between
exposures and the non-uniformity of the sampling.  To attempt to
quantify these uncertainties we create simulated lightcurves with
exactly the same sampling and integration times as the real data.  We
begin with a modelled ellipsoidal lightcurve, to ensure that the
double-sine approximation used does not provide an exact match and
that any systematic error introduced by the inadequacy of this model
is reproduced by the simulation.  This was calculated assuming $M_{\rm
X}=10$\,M$_{\odot}$, $q=0.067$, $i=40^{\circ}$, $R_{\rm
disc}=0.45$\,R$_{\rm L1}$.  These are arbitrary choices and are only
intended to produce a representative simulated lightcurve; this model
was not used to detrend the data.  With these parameters a reasonable
agreement was obtained with the observations, however.  To this was
added a model noise lightcurve, calculated using the method of Timmer
\& K\"{o}nig (1995).  Several models for the PDS were tested, broken
power-laws with a break close to that observed, $10^{-3}$\,Hz, at the
edge of the useful coverage, $10^{-4}$\,Hz and well outside the range
sampled, $10^{-5}$\,Hz.  In each case the power-law part of the
spectrum was matched to the observed slope and normalisation.
Lightcurves were created with 1\,s time resolution and binned up so
that high frequency variability would be aliased correctly.  We
included lower frequency variations than are well sampled by our
observations so that red noise leaks would have an effect.  Finally
Gaussian white noise was added using the errors derived from
photometry; for the comparison star these do represent the scatter in
the data accurately so the noise level should be comparable to the
real data.

\begin{figure}
\begin{center}
\hspace*{-8mm}
\epsfig{angle=90,width=3.8in,file=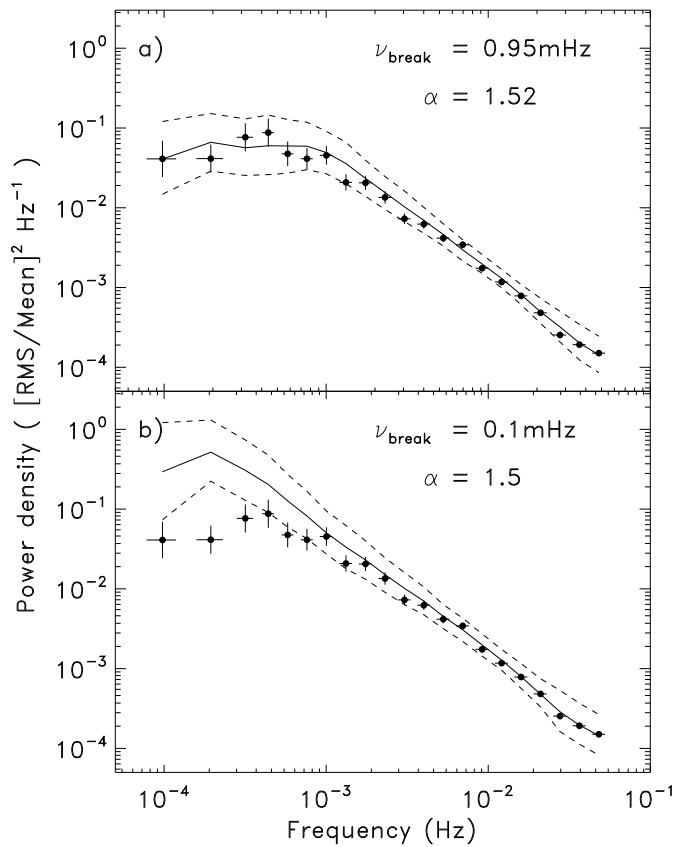}
\caption{Simulated PDS of \novamon\ compared with the data.
The solid line in each panel indicates the average of 1000 simulations.
The dashed lines indicate 1\,$\sigma$ confidence regions for
individual points.  a) Our best fit broken power-law model, indicating
that this is consistent with the data, and systematics due to
subtraction of the ellipsoidal modulation and red noise leaks do not
distort it substantially from the input model.  b) A case with
similar slope, but the break moved to the edge of the observable
range, $10^{-4}$\,Hz.  This is clearly a poor fit.}
\label{SimPDSFig}
\end{center}
\end{figure}

For each model of the PDS we calculate 1000 simulated lightcurves and
analyse them in exactly the same way as we did the real data.  We
create individual PDS with the same logarithmic frequency binning used
for the data, and then compare the range of PDS obtained with the
observations.  We find that if a broken power-law model fitting the
observations is input, then the simulated PDS do match the
observations (Fig.~\ref{SimPDSFig}a), i.e.\ we would not expect such a
PDS to be substantially distorted from the input form.  If the break
is dropped to $10^{-4}$\,Hz then the observed range of the PDS
corresponds to an unbroken power-law.  The output of the simulations
preserves this form fairly well, although there is some flattening at
low frequencies due to removal of some variability at the orbital
frequency and its first harmonic.  This flattening is at lower
frequencies than the observed break in the PDS of \novamon, and this
model does not agree with the data (Fig.~\ref{SimPDSFig}b).  Only 1
simulation out of 1000 produced sufficient flattening that all bins
below $10^{-3}$\,Hz have a power density below 0.1\,Hz$^{-1}$ as
observed.  We also ran simulations with even lower frequency breaks,
at $10^{-5}$\,Hz, to allow for the possibility of distortion by strong
red noise leaks.  These also provided a very poor agreement with the
observations.  We can therefore reject a model in which the intrinsic
PDS breaks at or below the edge of the observed range, and conclude
that the break in the PDS is probably real.

\subsection{X-ray Nova Mus 1991}

We calculated a PDS for \novamus\ in exactly the same way as for
\novamon.  It is also shown on Fig.~\ref{CompPDSFig}.  The PDS of the
raw lightcurve shows no sign of the break seen in \novamon, instead
possessing an unbroken power-law PDS.  The slope is very similar to
that of \novamon, however, $-1.56$.

Of course, just as subtracting a fitted ellipsoidal modulation from
\novamon\ may distort the PDS if some aperiodic power is removed as
well, so also an uncorrected ellipsoidal modulation could distort the
PDS of \novamus, although in the opposite sense.  To test for this, we
also tried subtracting the representative (not fitted) ellipsoidal
modulation shown in Fig.~\ref{NovaMusLCFig} from the lightcurve before
calculating the PDS.  Since we cannot be confident that this is an
accurate model, the resulting PDS may be no more correct that without
subtraction, but the differences between them should indicate the
sensitivity to the ellipsoidal contribution.  We see that a break does
emerge, but at a lower frequency than seen in \novamon.  Without being
sure of the ellipsoidal contribution we cannot claim that this break
is real, but we can say that any break in the PDS of \novamus\ must be
at $\la3\times10^{-4}$\,Hz.  Removing the ellipsoidal contribution
also flattens the PDS somewhat, as it otherwise introduces a red leak
across a range of frequencies which steepens the PDS.

\subsection{X-ray Nova Vel 1993}
We attempted to calculate a PDS for \novavel, but the combination of
low count rate and low variability amplitude meant that it was not
well-defined over a useful frequency range.  There is little prospect
of substantially improving on this, unless \novavel\ were caught in a
state showing a higher level of flaring.
%
%%%%%%%%%%%%%%%%%%%%%%%%%%%%%%%%%%%%%%%%%%%%%%%%%%%%%%%%%%%%%%%%%%%%%%%%%%%%%%%
%
%
\section{A serendipitous variable star discovery}

\begin{figure}
\begin{center}
\fbox{\epsfig{width=3.22in,file=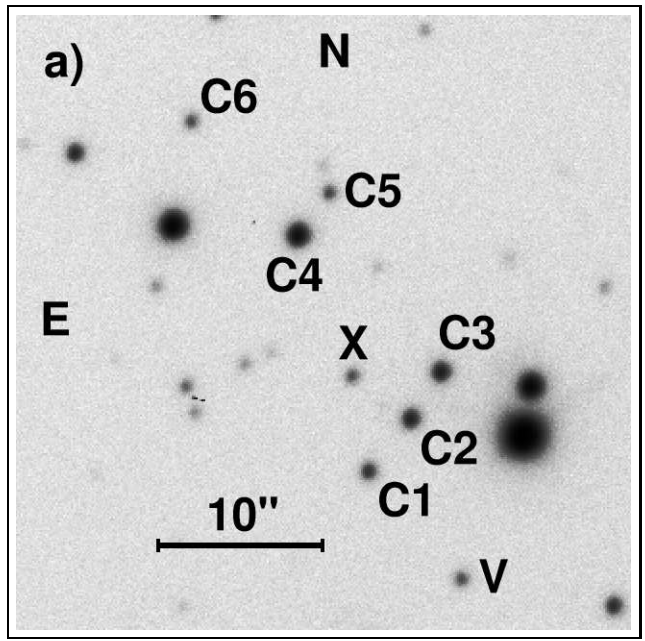}}\\
\hspace*{-8mm}
\epsfig{angle=90,width=3.8in,file=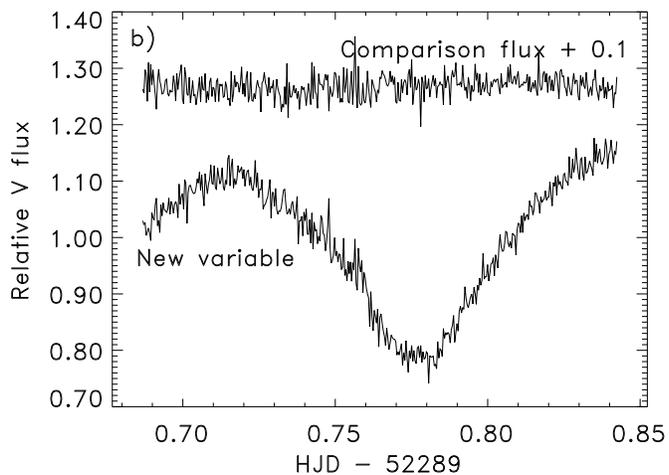}
\caption{a) The immediate field of \novamus, marked X, from one of our
best images.  The new variable is marked V.  Stars
C1 to C4 were combined as a reference for differential photometry of
both stars.  Stars C5 and C6 were used to check the extraction and
estimate errors; C6 is shown on lightcurves as it is closer to the
brightness of \novamus.  b) Lightcurve of the new variable star.  The
same comparison star as was used for \novamus, C6, has been shown;
this is somewhat brighter than the new variable.}
\label{NewVariableFig}
\end{center}
\end{figure}

One of the potential comparison stars we examined for \novamus\ was
revealed to be itself a variable.  This is about 14\,arcsec SSW of
\novamus, at RA $11^{\rm h}26^{\rm m}25\fs3\pm0\fs2$, Dec
$-68\degr40\arcmin44\farcs5\pm1\farcs0$ (J2000), based on
interpolation between 18 surrounding USNO A2.0 stars (Monet et al.\
1998).  From an approximate photometric calibration we estimate
$\left< V \right> = 20.5$, where the average is obviously over the
observed lightcurve, not a whole cycle.  It displays a smooth
modulation rather than erratic variability, reminiscent of W~UMa
contact binaries (Fig.~\ref{NewVariableFig}).  From the maximum and
minimum which are fully observed, a full amplitude of 0.37\,mag is
measured.  The later incomplete maximum, however, suggests an
asymmetric higher peak, and a full amplitude of $\ga0.40$\,mag.  Of
course, if the other minimum is deeper then the full amplitude will be
even larger.  The period of the modulation is either
$2.87\pm0.02$\,hrs if single peaked, or $5.73\pm0.03$\,hrs if double
peaked.  The latter seems more likely, given the unequal maxima
observed, and is a typical period for a W~UMa star (Maceroni \& van't
Veer 1996).  The asymmetric maxima are quite common to this class of
objects (e.g.\ Davidge \& Milone 1984 and references therein) and are
often attributed to star spots on these active late type binaries.
The difference we see, $\ga0.037$\,mag, is comparable to that seen in
other systems, and indeed the same difference as in the prototype,
W~UMa (e.g.\ Maceroni \& van't Veer 1996).  In short, it is most
likely that this variable is a W~UMa binary, or a related type.  A
more comprehensive study, based on several cycles of variability,
would be needed to draw a more confident conclusion.  Such a study
would likely be possible with existing images already obtained to
study ellipsoidal modulations in \novamus, as most such images should
also include the new variable.

%
%%%%%%%%%%%%%%%%%%%%%%%%%%%%%%%%%%%%%%%%%%%%%%%%%%%%%%%%%%%%%%%%%%%%%%%%%%%%%%%
%
\section{Discussion}
\label{DiscussionSection}
\subsection{What determines the variability amplitude?}
Zurita et al.\ (2002a) compared observed levels of variability with
various system parameters.  The only correlation they found was with
the binary inclination, but as they explain, this is somewhat
misleading.  This is because it actually arises from a more
fundamental correlation with the veiling; the fractional variability
correlates with the fractional contribution of disc light to the
total.  This is a very sensible result, but not trivial, since it
indicates that the variability is associated with the disc, as
expected, and not with activity on the companion star.

\begin{figure}
\hspace*{-8mm}
\epsfig{angle=90,width=3.8in,file=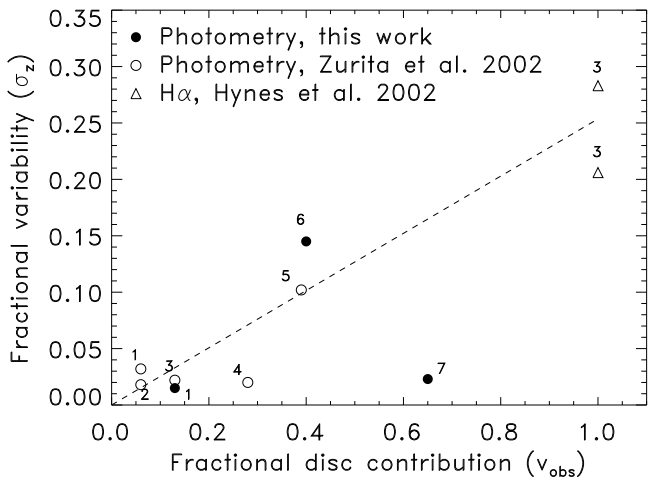}
\caption{Correlation between the fractional variability and the
fractional contribution of disc light.  Open circles are from Zurita
et al.\ (2002a), filled circles are from this work.  Triangles are
{\em spectroscopic} observations from Hynes et al.\ (2002).  The
sources are: 1. \novamon, 2. GS~2000+25, 3. V404~Cyg, 4. Cen~X-4,
5. GRO~J0422+32, 6. \novamus, 7. \novavel.  The dashed line is a
linear fit, passing through the origin, to all of the photometric
points except that of \novavel.}
\label{CorrelationFig}
\end{figure}

We illustrate the correlation using both our data and that of Zurita
et al.\ (2002a) in Fig.~\ref{CorrelationFig}.  We compare the
fractional variability in our detrended lightcurves ($\sigma_z$) with
the observed veiling fraction ($v_{\rm obs}$).  For \novamon\ and
\novavel\ we use the detrended lightcurves with the relative phasing
of the two sinusoids left free, but the difference between this and
the fixed phasing model was negligible.  For \novamus, the ellipsoidal
modulation is not adequately defined by the data so we do not detrend
the lightcurves.  Note that points from Zurita at al.\ (2002a), are
typically for a redder bandpass than our $V$ band measurements.  For
the veiling we use observed values based on spectroscopy,
interpolating where more than one wavelength is available.  The values
we use are as collated by Zurita et al.\ (2002a), together with Orosz
et al.\ (1996) and Casares et al.\ (1997) for \novamus, and Filippenko
et al.\ (1999) for \novavel.

A precise correlation between veiling and fractional variability would
obviously depend upon a number of assumptions and violation of these,
together with the uncertainty in the measurements, will introduce
scatter in the plot.  The comparison can only be crude anyway, as the
rms variability is not ideal for comparing datasets; it is effectively
an integration of the PDS, but the limits of the integration depend on
the length of the observation and the time-resolution, and hence will
vary from dataset to dataset.  The most serious physical assumption is
that the veiling source is the same as the flaring source.  This might
not be the case, for example, if the veiling came from the whole disc,
but the flaring only from the inner edge, or the stream-impact point.
If the flaring source does differ from the veiling source, it could
have a different spectrum, and hence it would be invalid to combine
$R$ and $V$ band observations in the same plot as we have done.  There
might also be differences in visibility, with, for example, the disc
being foreshortened, but emission from an inner spherical flow being
closer to isotropic.  This would lead to an inclination-dependent
scatter.  Finally intrinsic variations between sources will introduce
further scatter.  Consequently, it is not surprising that there is
significant scatter.  In spite of this, a correlation is visible;
certainly for objects where the veiling (i.e.\ the disc contribution)
is small, the variability amplitude is always small as expected.

It is interesting to further extend the comparison beyond optical
photometry.  Time resolved observations have also been performed in
H$\alpha$ for V404~Cyg (Hynes et al.\ 2002).  These observations are
useful because unlike the optical photometry, they should not show a
significant contribution from the companion star; the fractional disc
contribution is near unity.  We have therefore added the fractional
variability from these data to Fig.~\ref{CorrelationFig} for
comparison.  These points do lie relatively close to the line; they
certainly continue the sense of the extrapolation.  This is not
surprising given that Hynes et al.\ (2002) demonstrated that line and
continuum flares in V404~Cyg are correlated.  It is possible that {\em
all} optical variability in quiescent SXTs, continuum and line, has a
relatively uniform undiluted fractional variability of
20--30\,percent.  In fact, the {\it Chandra} X-ray observation of
V404~Cyg also showed a comparable level of variability (Kong et al.\
2002).

\novavel\ does appear to show much less variability than expected from
the very large veiling (60--70\,percent) estimated by Filippenko et
al.\ (1999).  Furthermore, their estimate was done at 6300\,\AA, and
we might expect a {\em higher} disc fraction in the $V$ band.  Such a
large value does, however, seem difficult to reconcile with the
pronounced ellipsoidal modulation which we observe.  It could be that
the disc was fainter at the time of our observations than when
Filippenko et al.\ (1999) observed it, and hence that the veiling is
less.  We should also consider, however, that the veiling estimate for
\novavel\ is probably the least certain of those in the sample.  It is
faint, and the existing spectra were heavily blended with the bright
star (more so than for our photometry).  The spectral type is not well
determined; Filippenko et al.\ (1999) use an M0 template to estimate
the veiling, but note that the spectral type could be as early as K6.
The heavy blending may also result in some contamination of the
spectrum of \novavel\ by the brighter star, reducing the validity of
the veiling determination further.  Consequently, we cannot be
confident that \novavel\ does show significantly less variability than
the trend suggested by the other sources in Fig.~\ref{CorrelationFig}.

\subsection{Lightcurves compared}
Since our Gemini observations of \novamon\ and \novamus\ show levels
of variability consistent with their veiling, and they have similar
orbital periods (7.8 and 10.4\,hrs respectively), we might expect that
the characteristics of the variability, such as the range of
timescales present, should be similar; the only major differences
between the observed variability properties should be in the amount of
dilution by the non-varying light from the companion star.  Some
difference is already suggested by the PDS shown in
Fig.~\ref{CompPDSFig}; below 1\,mHz, the PDS of \novamon\ appears to
flatten, while that of \novamus\ does not.  Above 1\,mHz, however, the
PDS are similar apart from the differing normalisations.  The PDS is
never the whole story, however, as phase information is not preserved.
We therefore should also directly compare the lightcurves.
Superficially, at least, these look very different.  We must be a
little careful, however; in \novamus, the ellipsoidal modulation is
weaker and the flaring much stronger than in \novamon.  The prolonged
flares seen in \novamus\ might, if scaled down in amplitude and
superposed on a strong ellipsoidal modulation, appear to be
undetectable or look like residuals to the fit to the modulation
rather than real flares.  Equally, the short, sharp flares in
\novamon\ would be less striking when observed at lower time
resolution.  A more realistic comparison is therefore to rebin the
\novamon\ lightcurve by a factor of 4 (giving a cycle time of about
32\,s, the same as used for \novamus), and scale the variations seen
in \novamus\ down to the same red-noise power and superpose them upon
an ellipsoidal modulation.

\begin{figure}
\hspace*{-8mm}
\epsfig{angle=90,width=3.8in,file=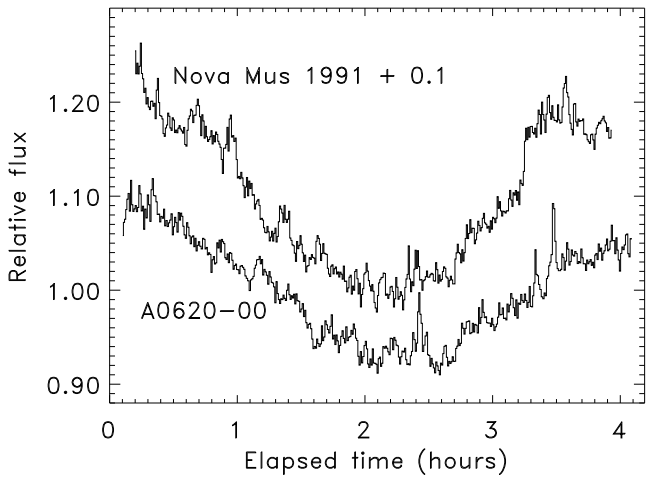}
\caption{A less biased comparison between the lightcurves of \novamon\
and \novamus.  The lightcurve of \novamon\ has been rebinned to the
same time resolution as that of \novamus.  The \novamus\ lightcurve
has been scaled down to what would be expected for the lower veiling
of \novamon, and an ellipsoidal model appropriate to \novamon\ has
been added.}
\label{LCCompFig}
\end{figure}

The results are shown in Fig.~\ref{LCCompFig}.  After applying this
processing to the lightcurve of \novamus, it does look more similar in
character to that of \novamon.  The drop at the beginning of the
lightcurve now blends indistinguishably with the synthetic ellipsoidal
modulation.  The strong, broad feature near the end is still visible,
though less dramatic.  At higher frequencies, however, one could
easily believe that the two lightcurves are of the same source, in the
same state.  This analysis therefore supports the assertion that some
of the very obvious differences between the lightcurves, at least of
sources of similar orbital period, do arise from the differing
amplitudes of the ellipsoidal and flaring components, which in turn
are simply dependent on the viewing geometry.  Differences in the low
frequency PDS do suggest some real differences in the variability
properties, however, and may give rise to some of the vertical scatter
in the correlation plot (Fig.~\ref{CorrelationFig}).  This does not
necessarily require that the origins of the variability in \novamon\
and \novamus\ differ; both PDS can be accommodated in a broken
power-law model, with the break out of the observable range in
\novamus.

\subsection{Is the variability the same as that in cataclysmic
variables?}
\label{CVSection}
Further clues to the origin of the variability may be obtained through
a comparison with cataclysmic variables (CVs).  We might expect the
outer regions of the disc to be rather similar in quiescent SXTs and
in quiescent dwarf novae (DNe).  Hence if the variability originates
from magnetic reconnection in the outer disc, or from the hot spot,
the variability properties should be similar.  Bruch (1992) has
compiled rapid photometry of many CVs.  If we select from his sample
only quiescent dwarf novae, a range of PDS slopes of $-1.6$ to $-2.6$
is seen; the best studied case is SS~Cyg which spans $-2.0$ to $-2.6$.
These are systematically steeper than we see, but not by so much as to
be conclusive; given only two well determined SXT PDS, we cannot rule
out the possibility that these represent the flattest examples drawn
from a similar distribution.  The variations in quiescent DNe have $B$
band full amplitudes of 0.18--1.26\,mag.  The latter is not directly
comparable to the rms we measure, but for a Gaussian distribution of
magnitudes would correspond to rms variations of 6--50\,percent; this
is a crude comparison, but at least enough to see that the amplitudes
observed are similar to those in SXTs (Fig.~\ref{CorrelationFig}).  In
DNe, as in quiescent SXTs, a range of amplitudes is expected due to
dilution of the variability by non-varying light, although unlike the
SXT case, this usually comes from the white dwarf, not from the
companion star.  Unfortunately, existing observations of quiescent DNe
such as those of Bruch (1992) do not adequately sample low-frequencies
to test for a break in the PDS, although a break has been seen in the
VY~Scl type CV, KR Aur during a high state (Kato, Ishioka \& Uemura
2002).  Thus the PDS of quiescent DNe and SXTs appear broadly similar,
although DNe may exhibit a somewhat steeper slope.  This comparison
is, however, inconclusive without a larger sample of well determined
PDS from both classes of objects.

\subsection{The significance of band-limited noise}
\label{PDSDiscussionSection}
\begin{figure}
\begin{center}
\hspace*{-8mm}
\epsfig{angle=90,width=3.8in,file=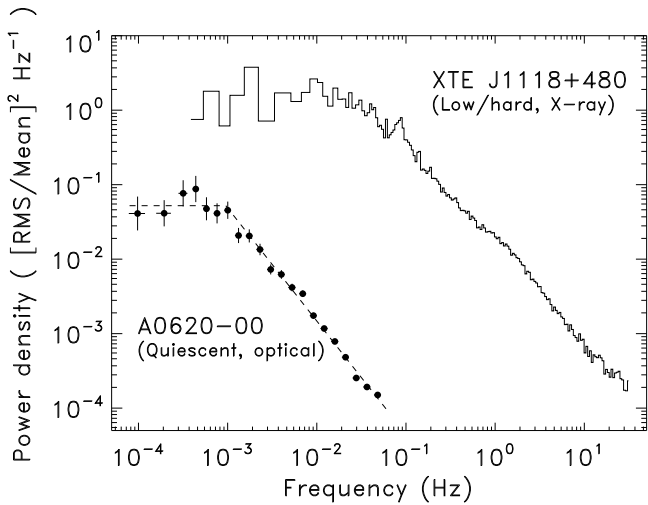}
\caption{PDS of \novamon\ compared with
XTE~J1118+480 in outburst (adapted from Hynes et al.\ in preparation).}
\label{XPDSFig}
\end{center}
\end{figure}

The presence of band-limited noise in \novamon, if not an artifact, is
intriguing.  This form of noise is also seen in low/hard state SXTs
(see Wijnands \& van der Klis 1999 and references therein).  This is
illustrated in Fig.~\ref{XPDSFig} by comparison with the X-ray PDS of
the low/hard state SXT XTE~J1118+480 (Hynes et al.\ in preparation).
The similarity of the PDS suggests that the optical variability could
have a similar origin and might be associated with the central X-ray
source, although we cannot say whether it represents direct emission
from an advective flow (e.g.\ self-absorbed synchrotron as predicted
by Narayan et al.\ 1996) or results from heating of the outer disc.
Like the quiescent state, the low/hard state is often interpreted as
involving a truncated accretion disc with an evaporated central
region.  An advective model has been successfully applied to
XTE~J1118+480 (Esin et al.\ 2001), and with spectral coverage into the
EUV it was clear in this case that the disc must be truncated (Hynes
et al.\ 2000; McClintock et al.\ 2001).  The most thorough analysis of
this dataset indicates an inner disc radius of $\sim350$\,R$_{\rm
Sch}$ (Chaty et al.\ 2002), significantly less than the
$10^3$--$10^5$\,R$_{\rm Sch}$ usually invoked for the quiescent state
(e.g.\ Narayan, Barret \& McClintock 1997).

There are differences in the two PDS shown in Fig.~\ref{XPDSFig}.  The
normalisation is clearly lower in the case of \novamon, as a fraction
of the mean flux, but this is easily understood as its optical light
includes a considerable contribution both from the companion star and
other parts of the accretion flow which exhibit only very low
frequency variability.  There does not appear to be an analogue of the
QPO seen in XTE~J1118+480, but this is not always present in the
low/hard state, either.  The most significant difference appears to be
in the break frequency, which is much lower in \novamon\ than in
XTE~J1118+480.  The origin of the break is not known, but it is
plausible to expect that it approximately scales with the size of the
inner region.  A low break frequency in quiescence would then be
expected, since the advective region is expected to be larger.  This
trend is supported by the much lower break frequency of
$4\times10^{-7}$\,Hz seen in the AGN NGC~3516 (Edelson \& Nandra
1999).  The simplest expectation would be that the break frequency
will scale linearly with the inner disc radius, but this may not
actually be the case; it will probably vary with some characteristic
length scale, which could depend on the mass transfer rate or other
parameters, as well as the inner disc radius.  Consequently there may
be additional factors of order unity.  If we assume that the
relationship is linear, and that the black hole masses in \novamon\
and XTE~J1118+480 are similar, then the ratio of break frequencies
implies an inner disc radius in \novamon\ of
$\sim1.1\times10^4$\,R$_{\rm Sch}$.  For comparison, advective models
of quiescent SXTs, and specifically of \novamon, assume an inner
radius of $10^3$--$10^4$\,R$_{\rm Sch}$ (Narayan et al.\ 1997).  Our
estimate is obviously very crude, and could be off by a factor of a
few, so the agreement with theoretical assumptions is reasonable.

The absence of a break in \novamus, or its lower frequency, however,
challenges this interpretation.  If no break is seen to a factor of
three lower in frequency then one would naively expect an inner disc
radius three times larger, $3\times10^4$\,R$_{\rm Sch}$.  This seems
large, at least for a relatively short period system such as \novamus.
Consequently, if this interpretation of the break is correct, it seems
likely that the break frequency does not scale linearly with radius,
and that other factors do come into play.  The absence of a clear
break in \novamus\ would make sense if the transition radius in
\novamon\ was unusually small at the time of observations.  Indeed,
the absence of a similar break in the PDS of Zurita et al.\ (2002a)
requires some variation in the PDS.  These considerations obviously
reduce the potential value of the break frequency for measuring the
transition radius, at least until the mechanism by which it arises is
understood.

One mechanism proposed for explaining the flaring in low/hard state
systems involves a cellular automaton model for an accretion disc (or
advective flow) in a self-organised critical state (Mineshige, Ouchi
\& Nishimori 1994a; Mineshige, Takeuchi \& Nishimori 1994b).  It is
assumed that material is injected into a region subject to an
instability if the density rises above a critical value; when this
occurs an avalanche is triggered and the energy release is manifest as
a flare.  This model can reproduce a band-limited noise PDS.  The
break frequency is related to the size of the unstable part of the
disc, which has subsequently been identified with an advective region
(Takeuchi \& Mineshige 1997).  Assuming a 10\,M$_{\rm odot}$ black
hole, a break frequency of $\sim10^{-3}$\,Hz implies the size of the
region is $\sim3000$\,R$_{\rm sch}$ (Mineshige et al.\ 1994a),
comparable to that obtained by scaling relative to XTE~J1118+480.
This radius also depends on the temperature and viscosity of the
region, so is quite uncertain.  Other aspects of the flare behaviour
are consistent with such a model, such as the distribution of flare
amplitudes and durations (Zurita et al.\ 2002a) and the roughly
symmetric flare profiles (Section~\ref{NovaMonLCSection}) as modelled
by Manmoto et al.\ (1996).

%%%%%%%%%%%%%%%%%%%%%%%%%%%%%%%%%%%%%%%%%%%%%%%%%%%%%%%%%%%%%%%%%%%%%%%%%%%%%%%
%
\section{Conclusions}
\label{ConclusionSection}
We have studied short-timescale flaring in three quiescent SXTs.  This
flaring is detected in all three sources, at low amplitudes in
\novamon\ and \novavel, and at a much higher level in \novamus.  With
the large aperture of Gemini it is possible to observe at higher time
resolution than in previous studies.  We find that the variability
extends to the shortest timescales observable, with pronounced changes
sometimes seen in 30\,s or less.  This is also indicated by the
extension of the red noise component in the PDS to 0.05\,Hz or even
higher.

A comparison of our observations with those of Zurita et al.\ (2002a)
supports their conclusion that the flares are associated with the
accretion flow rather than with the companion star.  This is clearly
shown by a correlation between the variability amplitude and the
fractional disc contribution to the spectrum.  The amplitude of
H$\alpha$ variations in V404~Cyg is also consistent with the
correlation, and it is likely that the line and continuum variations
have a related origin.

Compared to quiescent DNe, i.e.\ the nearest comparable systems not
containing a black hole or neutron star, our sources show similar
levels of variability, but perhaps with a somewhat flatter PDS.  A
rigorous comparison will, however, require a larger sample of objects
and more intensive observations.

In \novamon, we detect a low-frequency break in the PDS at
$\sim10^{-3}$\,Hz.  The PDS overall looks very similar to those of
low/hard state SXTs.  If the break frequency scales linearly with the
size of an inner evaporated region, then a comparison with
XTE~J1118+480 suggests that this region has size $\sim10^4$\,R$_{\rm
Sch}$ in \novamon\ in quiescence, although there are likely to be
other factors involved and this is an extremely crude estimate.  No
such break is confidently detected in \novamus; if present it must be
at $\la3\times10^{-4}$\,Hz.  It may be that the break frequency varies
from source to source and epoch to epoch, possibly in response to
changes in the inner truncation radius of the outer disc.  Again, more
intensive observations are needed to explore this behaviour.
%
%%%%%%%%%%%%%%%%%%%%%%%%%%%%%%%%%%%%%%%%%%%%%%%%%%%%%%%%%%%%%%%%%%%%%%%%%%%%%%%
%
\section*{Acknowledgements}
RIH would like to thank Phil Uttley for enlightening discussions on
power spectral analysis and the perils of red noise leaks, and Tom
Marsh for pointing out the similarity of the lightcurve of the new
variable to W~UMa stars.  Thanks also to Wei Cui for extracting the
lightcurve of XTE~J1118+480 used in Fig.~\ref{XPDSFig}, and to the
co-authors of Hynes et al.\ (in preparation) for permission to include
this in advance of publication.  This work is based on observations
obtained at the Gemini Observatory, which is operated by the
Association of Universities for Research in Astronomy, Inc., under a
cooperative agreement with the NSF on behalf of the Gemini
partnership: the National Science Foundation (United States), the
Particle Physics and Astronomy Research Council (United Kingdom), the
National Research Council (Canada), CONICYT (Chile), the Australian
Research Council (Australia), CNPq (Brazil) and CONICET (Argentina).
We would like to thank Claudia Winge for implementing the
observations, performing the pipeline reductions, and advice on the
data analysis.  RIH, PAC, and CAH acknowledge support from grant
F/00-180/A from the Leverhulme Trust.  RIH is currently supported by
NASA through Hubble Fellowship grant \#HF-01150.01-A awarded by the
Space Telescope Science Institute, which is operated by the
Association of Universities for Research in Astronomy, Inc., for NASA,
under contract NAS 5-26555.
%
%%%%%%%%%%%%%%%%%%%%%%%%%%%%%%%%%%%%%%%%%%%%%%%%%%%%%%%%%%%%%%%%%%%%%%%%%%%%%%%
%

%
\end{document}